\documentclass[a4paper, 11pt]{article}
\usepackage[margin=2.5cm]{geometry}
\usepackage{graphicx}
\usepackage{amsmath, amssymb}
\usepackage{mathtools}
\usepackage{braket}
\usepackage{authblk} 
\usepackage{xcolor}
\usepackage{placeins} 

\usepackage[patch=footnote]{microtype}
\usepackage[hidelinks]{hyperref}

\usepackage{algorithm}
\floatname{algorithm}{Protocol}

\usepackage{algpseudocode}
\usepackage{caption}
\usepackage{setspace}
\let\Algorithm\algorithm
\renewcommand\algorithm[1][]{\Algorithm[#1]\setstretch{1.2}}

\newcommand{\A}{\mathcal{A}}
\newcommand{\B}{\mathcal{B}}
\renewcommand{\P}{\mathcal{P}}
\newcommand{\T}{\mathcal{T}}
\newcommand{\CNOT}{\textrm{CNOT}}

\title{Cryptanalysis of four arbitrated quantum signature schemes}
\author[a]{Pierre-Alain Jacqmin\thanks{Email: pierre-alain.jacqmin@mil.be; ORCID: 0000-0001-9834-1433}}
\author[a]{Jean Li\'enardy\thanks{Email: jean.lienardy@mil.be; ORCID: 0009-0002-6377-0734}}
\affil[a]{\small{\textit{Department of Mathematics, Royal Military Academy, Brussels, Belgium}}}
\date{\today}

\begin{document}

\maketitle

\abstract{Arbitrated quantum signature (AQS) schemes aim at ensuring the authenticity of a message with the help of an arbitrator. Moreover, they aim at preventing repudiation, both from a sender that denies the origin of a message, and from a receiver who disavows its reception. Such protocols use quantum communication and are often designed to protect quantum messages. In this paper, we study four recently submitted AQS schemes and propose attacks on their security.

Firstly, we look at Zhang, Sun, Zhang and Jia's AQS scheme which aims at signing quantum messages with chained CNOT encryption. We show that the sender can repudiate her messages and make false allegation of reception. Moreover, we show that a dishonest receiver can forge signatures.

Secondly, we analyse Ding, Xin, Yang and Sang's AQS protocol to sign classical messages based on GHZ states. We show that both the sender and the receiver have simple repudiation strategies.

Thirdly, we study Lu, Li, Yu and Han's AQS scheme that uses controlled teleportation to protect quantum messages. We expose forgeries, false allegation attacks and the possibility of repudiation by both parties.

Fourthly, we focus on the AQS scheme by Zhang, Xin, Sun, Li and Li designed to sign classical messages without entangled states. We show that one can disavow the reception of messages, and that information-theoretic security is not achieved for other security goals.}

\medskip

{\small\textbf{Keywords:} arbitrated quantum signature, cryptanalysis, quantum message, authentication, non-repudiation, forgery.}

\tableofcontents
\listofalgorithms

\section{Introduction}

Signatures are one of the most important primitives of cryptography, as they allow binding a message to its originator while ensuring integrity and non-repudiation. Classical schemes achieve this with public-key primitives, but their security typically rests on computational assumptions that may be threatened by quantum algorithms. Although post-quantum cryptographic signatures have been proposed as a solution to this problem, this has, in parallel, motivated signature mechanisms whose security derives from physics rather than hardness assumptions.

It is well known that public-key signatures of quantum messages with unconditional security are impossible~\cite{BCGST02}. The seminal work of Zeng and Keitel~\cite{ZK02} paved the way for the field of arbitrated quantum signatures (AQS), which aims to overcome this limitation by means of an arbitrator, in addition to the sender and the receiver.

The goal of AQS is often to sign a generic quantum state opposed, say, to signing a classical message by quantum means, as done in quantum digital signatures~\cite{GC01}. This is performed by involving an arbitrator in the protocol, who participates to the signing process and whose role is to resolve disputes. There are multiple models of arbitrators: it can be fully trusted, honest-but-curious (following the protocol but trying to read the message) or adversarial (trying to forge a signature).

The security goals of AQS are unforgeability and non-repudiation. The former is the impossibility for anyone other than the signer to create a legitimate signature on their name, while the later, also referred to as impossibility to disavow, splits in non-repudiation of origin (the signer cannot claim that they did not sign the message) and non-repudiation of receipt (the receiver cannot claim that they did not receive it). With classical public-key digital signatures, usually only unforgeability and non-repudiation of origin are aimed. 
Non-repudiation of receipt is not a goal of classical signatures, and it is usually not achieved by signatures alone.
There exist classical protocols aiming to provide both non-repudiation of origin and receipt, typically referred to as fair non-repudiation protocols. They usually require additional assumptions such as trusted third parties, synchronization, or constraints on computing power~\cite{KMZ02}. In order to perform an AQS protocol, some secret is usually pre-shared between parties, such as a classical bit string or in the form of an entangled quantum state. Note that multiple copies of the quantum message are often required to perform the signature.

After the seminal work of Zeng and Keitel~\cite{ZK02,CL08,Zeng08}, in which the three parties pre-share a GHZ state, numerous modifications and improvements have been proposed. These include relying on Bell states instead of GHZ triplets~\cite{LCL09}, removing the need for entanglement~\cite{ZQ10}, modifying the encryption algorithm~\cite{ZQSSS13,ZZL13,ZQ13,ZLS14,KCJL15}, preventing denial-of-service attacks using Bell states~\cite{WLS14}, using chaos-based encryption~\cite{WXG14}, introducing decoy states~\cite{LQS14}, and chaining CNOT operations for encryption~\cite{LS15,ZSZJ17}.

In parallel, a substantial body of cryptanalysis has highlighted flaws in many published schemes. For example, \cite{ZQ10} showed that the initial schemes~\cite{ZK02,LCL09} are vulnerable to disavowal by Bob, who can claim that a legitimate signature is invalid. The works~\cite{CCH11,GQGW11} further demonstrated that the use of the qubit-wise quantum one-time pad (QOTP) makes the transmitted message malleable, enabling existential forgery in earlier protocols. The authors of~\cite{GQGW11} also noted that modification by Alice of messages between Bob and the arbitrator can lead to repudiation by Alice. This line of cryptanalysis continued with \cite{SDWLL11,LZC14,WLS14,LH15,ZSZWC17}, each targeting one or multiple schemes and paving the way towards more secure designs.

\paragraph{Contributions.} This article continues that cryptanalytic agenda by focussing on four recent schemes \cite{ZSZJ17,DXYS22,LLYH22,ZXSLL24} that, to our knowledge, have up to now only been partially analysed.

First, we focus on the scheme of~\cite{ZSZJ17} built upon the idea of key-controlled chained CNOT encryption, whose goal is to thwart forgery attacks caused by the malleability of the QOTP. It uses an encryption method that involves and mixes multiple registers of the quantum message as proposed in~\cite{ZZL13}.

Second, we analyse the scheme from~\cite{DXYS22} that relies on shared GHZ states and which aims at unforgeability and non-repudiation. A particularity of this scheme is that it does not rely on quantum teleportation, even though entangled states are shared.

In third place, we consider an AQS protocol based on controlled teleportation, designed by Lu et al.~\cite{LLYH22}. There, a $5$-qubit quantum state shared between the parties serves as a medium to perform quantum teleportation. 

Finally, we address an AQS scheme that signs classical messages without relying on entangled states~\cite{ZXSLL24}, thereby aiming at better performance and simpler implementation.

The outline of this article is straightforward: in each of Sections~\ref{sec:scheme_CNOT_1}--\ref{sec:scheme_no_entangled}, a scheme is first described. If there are any ambiguities in the protocol description, they are raised and discussed. Security claims of the protocol are then stated, analysed, and challenged. For each protocol, we show that most security claims are not met, either partially or completely. In the last section, we draw short conclusions. In the rest of the present section, we present the general structure of an AQS protocol, which is shared by all the schemes reviewed in this article, and we fix some notation that we will later use.

\paragraph{Generic description of an arbitrated quantum signature scheme.} AQS protocols are usually divided into three procedures, although we consider two additional procedures in this paper for better precision. These procedures involve a sender Alice~($\A$), a receiver Bob~($\B$) and an arbitrator Trent~($\T$). A first initialisation step, that we note \textsc{Init}, is usually required to exchange secret bit strings or pre-share quantum states if required by the protocol. This step is assumed to be repeated for each new message to be signed, except if stated explicitly. The actual signature of the message is done during the \textsc{Sign} procedure, which takes as input one or multiple copies of a (classical or quantum) message, as well as the data established during the \textsc{Init} procedure, in order to produce a quantum signature. This procedure often ends with the signature being sent to the receiver Bob. The third step of the protocol is the verification of the signature, noted \textsc{Verify}, during which Bob gets confirmation that the message indeed comes from Alice and has not been modified. Two additional procedures are needed for dispute resolution. Although they are not part of the signature protocol itself, they are nevertheless required in the case where parties disagree at some point after the initial protocol. One of them is the proof of origin, noted \textsc{Proof-of-Origin}, in which Bob proves to the arbitrator that Alice has indeed sent him a certain message. The second one is the proof of receipt, noted \textsc{Proof-of-Receipt}, in which Alice proves to the arbitrator that Bob has indeed received a certain message from her. These additional procedures must be precisely formulated, as the security claims (non-repudiation of origin and non-repudiation of receipt) will depend on how they are performed. We stress that for certain protocols, these procedures are not supposed to be run several times for the same message to sign. For instance, their second run might even be badly defined because a quantum state has been sent or measured.

The communication during the procedures \textsc{Init}, \textsc{Proof-of-Origin} and \textsc{Proof-of-Receipt} is assumed to be made with authenticity of the originator, message integrity, and guaranteed reception, contrarily to \textsc{Sign} and \textsc{Verify}. Moreover, the communication during \textsc{Init} is assumed to be secret. Some protocols require a public board, that we note~$\P_b$, on which parties can write public classical messages. It is usually assumed that communications with the public board are authenticated and that the public board is always available.

As far as security properties are concerned, we require that an arbitrator gives a judgement in favour of the complainant Bob for a \textsc{Proof-of-Origin} procedure on a certain message if and only if Alice indeed sent this message. If Bob manages that Trent gives a judgement in his favour when Alice did not sent it, we call it a \emph{forgery} attack; while if Alice manages that Trent gives a judgement in her favour when she has indeed sent the message, it is a \emph{repudiation of origin} attack. In addition, we require that an arbitrator gives a judgement in favour of the complainant Alice in a \textsc{Proof-of-Receipt} procedure on a certain message if and only if Bob indeed received this message. If Bob manages to make Trent judge in his favour when his has actually received the message, we call it a \emph{repudiation of receipt} attack or a \emph{disavowal of reception}; while if Alice makes Trent say that Bob received a message that he did not, we call it a \emph{false allegation} attack. Avoiding false allegations from Alice is often not considered as a security goal. However, it should not be omitted; otherwise, a trivial \textsc{Proof-of-Receipt} procedure could consist of Trent always being in favour of Alice. This would avoid any repudiation from Bob, but would not give any meaningful guarantees.

\paragraph{Notation.} We denote a classical bit string by a lowercase letter (e.g., keys~$k$), and its single bits by subscripts $k=k_1\dots k_n$. Its bit-length is noted with~$|\cdot|$ (e.g., $|k|=n$). We use the standard bra-ket notation for quantum states. The tensor product of states is often implicit, that is, we write $\ket{X}\ket{Y}$ for $\ket{X}\otimes \ket{Y}$. We even use the notation $\ket{X,Y}$ when no confusion is possible. Note that the respective lengths of the registers of $\ket{X}$ and $\ket{Y}$ are assumed known by legitimate parties. Qubits are two-dimensional quantum states that are represented by unit vectors in~$\mathbb{C}^2$. Given a state defined as a tensor product of $n$ qubits, we call $n$ its length.

The four Pauli matrices are $\{\sigma_\alpha\}_{\alpha=0}^3=\{I,X,Y,Z\}$ and the Hadamard operator is noted~$H$. We recall that each of these operators is an involution $(\sigma_\alpha)^2=I=H^2$ for any~$\alpha$, that different non-identity Pauli matrices anticommute ($XY=-YX$, $XZ=-ZX$ and $YZ=-ZY$) and that Pauli and Hadamard operators satisfy:
\[ HX=ZH, \quad HZ=XH,\quad HY = -YH.\]
Hadamard and Pauli gates act qubit-wise: given an $n$-qubit register $\ket{M}=\sum_j c_j\bigotimes^n_{i=1} \ket{M^j_i}$ and a bit string $y$ of length greater than or equal to~$n$, $\ket{S} = H^y\ket{M}$ means that $\ket{S}=\sum_j c_j\bigotimes^n_{i=1} \ket{S^j_i}$ where $\ket{S^j_i} = H^{y_i}\ket{M^j_i}$ and analogously for Pauli gates. 
We use the standard computational basis $\{\ket{0}, \ket{1}\}$ composed of eigenvectors of $Z$ and the diagonal basis $\{\ket{+},\ket{-}\}$, defined as $\ket{\pm}=\frac{1}{\sqrt{2}}(\ket{0}\pm\ket{1})$. Given a classical message~$m$, we note $\ket{m}$ its encoding in the computational basis: $\ket{m}=\bigotimes_i\ket{m_i}$ with $\ket{m_i}=X^{m_i}\ket{0}$. In other words, we apply the canonical mapping $m_i\mapsto \ket{m_i}$ to $m_i\in\{0,1\}$.

When a party transmits a message, this is indicated with the symbol~$\to$. The notation $\A\leftrightarrow\B:m$, used during \textsc{Init} procedures, indicates that the classical message $m$ has been exchanged between $\A$ and~$\B$.

In a description of a protocol, the phrase ``$\mathcal{X}$: recover $\ket{\cdot}$" is meant to apply the appropriate operations on a register in $\mathcal{X}$'s possession using information available to $\mathcal{X}$ in order to recover a given state, for example, as in standard quantum teleportation or to decrypt an encrypted message.

In protocols, we sometimes use a semicolon to separate instructions. We also use the notation $X=Y$~?~$P_T : P_F$ to indicate that an equality test is performed between $X$ and~$Y$. If the test succeeds, that is, if one evaluates $X=Y$, then the instructions $P_T$ are executed, else the instructions $P_F$ are executed. Many of these tests are of the form $E(Z)=Y$~?~$P_T : P_F$, that is, with $X=E(Z)$ being the result of an invertible function~$E$. When this is the case, we assume that if the test is passed, the party implicitly recomputes the input states $Z$ and~$Y$. Note that if $X$ and $Y$ are quantum states, one usually needs several copies of them to trustfully perform such an equality test by using SWAP tests~\cite{BCWdW01}.

Some instructions are written in bold and indicate that a party aborts: the party does not continue the procedure. When such an instruction is reached, it is assumed that all other parties are aware of it. Specifically, in a $\textsc{Verify}$ procedure, the instruction \textbf{abort} simply indicates that the party aborts the procedure, \textbf{accept} means that the receiver considers the signature as valid and ends the procedure, \textbf{reject} indicates that the receiver considers the signature invalid and aborts. In a phase of dispute resolution, the instructions \textbf{favour $\A$} and \textbf{favour $\B$} respectively mean that the arbitrator gives a judgement in favour of Alice or Bob, and the procedure is stopped.

\paragraph{Acknowledgment.} Both authors were funded by the Defense Funded Research (DFR) study DAP/23-01.

%

\section{Cryptanalysis of a scheme based on chained CNOT}
\label{sec:scheme_CNOT_1}

In this section, we analyse the AQS scheme proposed by Zhang et al.~\cite{ZSZJ17}, aimed at the signature of quantum messages and based on the idea of chained CNOT encryption~\cite{LS15,LH15}. We first describe the scheme and its building blocks before analysing its security claims.

\subsection{Description of the protocol}
\label{subsec:scheme_CNOT_1:description}

\begin{algorithm}[p!]
    \begin{algorithmic}
\Procedure{Init}{}
\State 1. $\A\leftrightarrow\B$: $k_{AB}$.
\State 2. $\A\leftrightarrow\T$: $k_{AT}$.
\State 3. $\B\leftrightarrow\T$: $k_{BT}$.
\EndProcedure{}
\vspace{1mm}

\Procedure{Sign }{\text{on 3 copies $\ket{M^0}=\ket{M^1}=\ket{M^2}$ of} $\ket{M}=\bigotimes_{i=1}^n\ket{M_i}$}
\State 1. $\A$: choose random~$r$; $\ket{P^i}=E^{\text{KCCC}}_{r}\ket{M^i}$ for $i\in\{0,1,2\}$.
\State 2. $\A$: $\ket{P}=\ket{P^0}$;
\quad $\ket{R_{AB}}=E^{\text{KCCC}}_{k_{AB}}\ket{P^1}$;
\quad $\ket{S_A}=E^{\text{KCCC}}_{k_{AT}}\ket{P^2}$.
\State 3. $\A$: $E^{\text{KCCC}}_{k_{AB}}\big(\ket{P},\ket{R_{AB}},\ket{S_A}\big) \to \B$.
\EndProcedure
\vspace{1mm}

\Procedure{Verify}{}
\State 1. $\B$: recover $\ket{P},\ket{R_{AB}},\ket{S_A}$.
\State 2. $\B$: $\ket{Y_{BT}}=E^{\text{KCCC}}_{k_{BT}} \big(\ket{P}, \ket{S_A}\big)\to \T$.
\State 3. $\T$: recover $\ket{P},\ket{S_A}$.
\State 4. $\T$: $E^{\text{KCCC}}_{k_{AT}}\ket{P} = \ket{S_A} \ ?\ v_T= 1 : v_T=0$; $v_T\to \P_b$.
\State 5. $\B$: $v_T=0$ ? \textbf{reject} : \text{continue}.
\State 6. $\T$: $\ket{Y_{TB}} = E^{\text{KCCC}}_{k_{BT}} \big(\ket{P}, \ket{S_A}\big)\to \B$.
\State 7. $\B$: recover $\ket{P}, \ket{S_A}$.
\State 8. $\B$: $E^{\text{KCCC}}_{k_{AB}}\ket{P} = \ket{R_{AB}} \ ? \ v_B= 1 : v_B=0$; $v_B\to \P_b$.
\State 9. $\B$: $v_B=0$ ? \textbf{reject} : \text{continue}.
\State 10. $\A$: $v_T=v_B=1 \ ?\ r\to \P_b$ : \textbf{abort}.
\State 11. $\B$: recover $\ket{M^0}$ from $\ket{P}$ and~$r$.
\State 12. $\B$: store signature $(\ket{M^0},\ket{S_A},r)$; \textbf{accept}.
\EndProcedure

\Procedure{Proof-of-Origin }{\text{on input} $\ket{\overline{M}}=\bigotimes_{i=1}^n\ket{\overline{M}_i}$}
\State 1. $\T$: $v_T=v_B=1$ ? continue : \textbf{favour~$\A$}.
\State 2. $\B$: $\ket{S_A} \to \T$.
\State 3. $\T$: $\ket{S_A}=E^{\text{KCCC}}_{k_{AT}} E^{\text{KCCC}}_{r}\ket{\overline{M}}$ ? \textbf{favour~$\B$} : \textbf{favour~$\A$}.
\EndProcedure

\Procedure{Proof-of-Receipt }{\text{on input} $\ket{\overline{M}}=\bigotimes_{i=1}^n\ket{\overline{M}_i}$}
\State 1. $\T$: $v_T=v_B=1$ ? \textbf{favour~$\A$} : \textbf{favour~$\B$}.
\EndProcedure
\end{algorithmic}

\caption[AQS Protocol~\ref{alg:Algo_CNOT}, of Zhang, Sun, Zhang and Jia]{AQS protocol of~\cite{ZSZJ17} (the encryption of $\ket{M}$ to $\ket{P}$ is chosen as $E^{\text{KCCC}}$).}
\label{alg:Algo_CNOT}
\end{algorithm}

The main building block for the security of this scheme is a multi-layer encryption technique aimed at diffusing the information in each qubit of the message across multiple qubits, rather than merely performing a qubit-wise operation. The entire protocol is given as Protocol~\ref{alg:Algo_CNOT}.

The first encryption layer is performed by using multiple CNOT gates in a chain-like fashion. We recall that the gate $\CNOT_{i,j}$ (with $1\leqslant i,j\leqslant n$, $i\neq j$) applied on an $n$-qubit state $\ket{P}=\bigotimes_{\ell=1}^n \ket{P_\ell}$ with $P_1,\dots,P_n\in\{0,1\}$ acts on the $i^{\text{th}}$ and $j^{\text{th}}$ tensor register, mapping $\ket{P_i}\ket{P_j}$ given in the computational basis to $\ket{P_i}\ket{P_i\oplus P_j}$ (and this definition is extended linearly). Furthermore, $\CNOT_{i,i}$ is the identity operator. The chained $\CNOT$ encryption of an $n$-qubit message $\ket{P}$ is noted $E_{k^1}^{\CNOT}\ket{P}$ and requires a key $k^1=(k^1_{(1)},\dots,k^1_{(n)})$ which is a permutation of $(1,\dots,n)$ where each $k^1_{(i)}$ is the image of~$i$. The permutation can be encoded in $\lceil\log_2(n!)\rceil$ bits, which we consider as the key length. The chained $\CNOT$ encryption is given by 
\[
E^{\CNOT}_{k^1} \ket{P} = \CNOT_{n,k^1_{(n)}}\cdots \CNOT_{2,k^1_{(2)}} \CNOT_{1,k^1_{(1)}} \ket{P}.
\]

The second encryption layer is~$H^{k^2}$, which consists in applying $H$ conditionally to the key bits of~$k^2\in\{0,1\}^n$.

The third and final step for the encryption is a permutation that depends on the key $k^3\in\{0,1\}^n$, that we denote by $E^{\text{perm}}_{k^3}$. The authors describe two such key-controlled permutations that can be applied to the qubits of a state.

The first one is a controlled pairwise transposition of the qubits $i$ and $n+1-i$. For each $i\in\{1,\dots,\lfloor\frac{n}{2}\rfloor\}$, let $\tau_i=k^3_i\oplus k^3_{n+1-i}$. Then, $S_{i,n+1-i}^{\tau_i}$ is a linear operator that acts on $\ket{P}=\bigotimes_{\ell=1}^n \ket{P_\ell}$ with $P_1,\dots,P_n\in\{0,1\}$ as the identity if $\tau_i=0$ and by swapping the qubits $\ket{P_i}$ and $\ket{P_{n+1-i}}$ if $\tau_i=1$. This action is extended on generic states by linearity. One then defines $E^{\text{transp}}_{k^3}=S_{\lfloor\frac{n}{2}\rfloor,\lceil\frac{n}{2}\rceil+1}^{\tau_{\lfloor\frac{n}{2}\rfloor}} \circ\cdots\circ S_{2,n-1}^{\tau_2} \circ S_{1,n}^{\tau_1}$. 

The second controlled permutation, denoted $E^{\text{rot}}_{k^3}=S_n^{(\tau)}$, is a rotation of the $n$ qubits. It rotates the qubits by $\tau$ positions where $\tau = (\sum_i k^3_i)\pmod{n}$:
\[
S_n^{(\tau)} \ket{P_1,P_2,\cdots,P_n} = \ket{P_{\tau+1},\cdots,P_n,P_1,\cdots,P_\tau}
\]
for $P_1,\dots,P_n\in\{0,1\}$, extended linearly.

The full encryption of a state $\ket{P}$ of length $n$ is denoted $E^{\text{KCCC}}_k\ket{P}$. It requires a key $k=k^1\|k^2\|k^3$ with $k^1$ a permutation of $(1,\dots, n)$ and $k^2,k^3\in\{0,1\}^n$:
\[
E^{\text{KCCC}}_{k^1||k^2||k^3} \ket{P} = E^{\text{perm}}_{k^3}\circ H^{k^2}\circ E^{\CNOT}_{k^1} \ket{P},
\]
where ``$\text{perm}$" is either ``$\text{transp}$" or ``$\text{rot}$", depending on the chosen controlled permutation, which is a public information chosen before the start of the protocol. The protocol is given\footnote{with a small correction to the protocol given in~\cite{ZSZJ17}. Indeed, the condition ``$V_T=0$" in (V2) appears to be an error. Consistency with the subsequent step requires $V_T=1$. We adopted this corrected condition in our analysis.} in Protocol~\ref{alg:Algo_CNOT}. We note that it does not use any pre-shared quantum states, such as Bell states or GHZ triplets.

\paragraph{Ambiguities.} We note that the description of~\cite{ZSZJ17} contains some ambiguities. We list and explain how we deal with these ambiguities in our analysis of the scheme.

A first ambiguity concerns the subkey sizes. Given a key $k=k^1\|k^2\|k^3$ used to encrypt a state of $n$ qubits, the key~$k^3$ (noted $L$ in~\cite{ZSZJ17}) is obtained from $k^2$ through a hash function: $k^3=f(k^2)$. However, the hash function has input size~$2n$, while $k^2$ is of length~$n$. We instead treat $k^3$ as independent from~$k^2$, both of length~$n$. This choice clears the ambiguity and furthermore removes the need for a hash function, whose presence could affect the ITS nature of the whole scheme, and which was only marginally useful, as the key length is dominated by that of $k^1$ for large~$n$. Moreover, this does not change the success probability of our attacks if the hash function is modelled as a random oracle.

Still regarding the subkey size, we note a suboptimal usage of the key~$k^3$. If the permutation chosen is $E^{\text{transp}}$, then only $\lfloor \frac{n}{2}\rfloor$ bits are required to specify the transposition. These are the bits denoted by $\tau_i$ above. For $E^{\text{rot}}$, an integer $\tau\in [0,n-1]$ is sufficient, hence requiring $\lceil \log_2(n)\rceil$ bits. In both cases, some of the $n$ bits of $k^3$ are superfluous. Moreover, while the distribution of the $\tau_i$ used by $E^{\text{transp}}$ is uniform if the $k_i^3$ are independent and uniformly distributed, the distribution of $\tau=(\sum_i k_i^3)\mod n$, used for $E^{\text{rot}}$, is not uniform for a uniform distribution of~$k^3_i$. While we do not know if this behaviour was desired, we decided to keep, in the following, the initial construction using the $n$ bits of~$k^3$. We stress that our attacks would still hold (although with smaller or equal probability) if $\tau$ was uniformly selected in $[0,n-1]$. 

A second imprecision regards the size of the whole keys and their re-use. The authors mention a single key length, which is equal to~$n$. However, as stated above, a key that encrypts a state of $n$ qubits must have a size of at least $2n+\lceil \log_2(n!)\rceil$ (or still $n+\lceil \log_2(n!)\rceil$ if we keep the hash function, and $n+\lceil \log_2(n!)\rceil+\lfloor\frac{n}{2} \rfloor$ or $n+\lceil \log_2(n!)\rceil+\lceil\log_2(n)\rceil$ if we had chosen optimal subkeys for $E^{\text{perm}}$). Furthermore, while the key $k_{AT}$ does encrypt and decrypt states of size~$n$, the key $k_{BT}$ is used for states of size~$2n$ (see steps~2 and~6 in the verifying procedure), hence it is of minimal length $4n+\lceil \log_2((2n)!)\rceil$. As for the key~$k_{AB}$, it is used to encrypt both a state of size $n$ and a state of size $3n$ in steps~2 and~3 of the signing procedure, respectively. It is unclear what the impact of reusing a key is. In the following, we suppose that two different keys $k_{AB}$ are used to prevent any interference. The size of the pre-shared secret between Alice and Bob is thus $8n+\lceil \log_2((3n)!)\rceil+\lceil \log_2(n!)\rceil$.

A third ambiguity is the transformation of $\ket{M}$ into $\ket{P}$ using~$r$, which is not explicitly stated. In the following, we assume that this is done using the same layered encryption $E^{\text{KCCC}}$, using a classical key $r$ of appropriate size.

A fourth ambiguity concerns the procedures \textsc{Proof-of-Origin} and \textsc{Proof-of-Receipt} since they are not explicitly described in the paper~\cite{ZSZJ17}. The \textsc{Proof-of-Origin}, where Bob wants to prove that Alice sent him a certain message~$\ket{\overline{M}}$, is quite intuitive, and the one we propose in Protocol~\ref{alg:Algo_CNOT} is probably what the authors have in mind. However, the procedure \textsc{Proof-of-Receipt}, in which Alice wants to prove that Bob received a certain message $\ket{\overline{M}}$ from her, is far less straightforward. According to our understanding of Section~4.2 of~\cite{ZSZJ17}, it works as follows: on input $\ket{\overline{M}}$ from Alice, Trent gives a judgement in favour of Alice if and only if $v_T=v_B=1$. 
In this case, Alice has a trivial false allegation attack: as the procedure does not depend on $\ket{\overline{M}}$, Alice can pretend to have sent any message to Bob. Although being trivial, we choose to describe this option in Protocol~\ref{alg:Algo_CNOT}.
We thought of another option, in which Trent asks Bob to send him the state $\ket{S_A}$ from the verification procedure to make an additional equality test, as in the \textsc{Proof-of-Origin} procedure. Since Bob could modify $\ket{S_A}$ as he wishes, he can trivially deny having received the message $\ket{\overline{M}}$ from Alice. 

A fifth issue concerns the set of messages Alice can sign. In the paper~\cite{ZSZJ17}, it is mentioned that she can only sign $n$-qubit messages of the factorised form $\ket{M}=\bigotimes_{i=1}^n\ket{M_i}$, although there exists $n$-qubit messages which are not of this form. The utility of this restriction is unclear to us as the scheme could easily be extended to any entangled state. Moreover, as no checks on format are made by the participants, we do not know what happens if an non-factorised message is used. Nevertheless, our attacks are not impacted by this restriction.

\subsection{Analysis of the security claims}

The authors of~\cite{ZSZJ17} base their security analysis on the fact that an error in the ciphertext may propagate with the decryption of $E^{\text{KCCC}}$ through different qubits in the plaintext. In this subsection, we show that despite the enhanced encryption used, there is still the possibility for Alice to repudiate a message and to make false allegations, and for Bob to do a forgery. This contradicts the claims of~\cite{ZSZJ17}.

\paragraph{Repudiation by Alice.} Suppose that there are two distinct quantum messages $\ket{M}$ and~$\ket{M'}$, with their corresponding states, as defined at step~2 of the \textsc{Sign} procedure, being~$\ket{P}$, $\ket{S_A}$, $\ket{R_{AB}}$ and $\ket{P'}$, $\ket{S_A'}$, $\ket{R_{AB}'}$. In the following, we suppose that Alice has sent $E^{\text{KCCC}}_{k_{AB}}\left(\ket{P'},\ket{R_{AB}},\ket{S'_A}\right)$ to Bob during the last step of the \textsc{Sign} procedure. This message is considered valid by the arbitrator during the \textsc{Verify} procedure and $v_T=1$. Alice wishes to intercept and modify the message $\ket{Y_{TB}}$ on its way back to Bob in step~6 so that he obtains $\ket{P}\ket{S'_A}$ instead of $\ket{P'}\ket{S'_A}$. If she succeeds, Bob's checks will pass, and he will consider the signature $(\ket{M},\ket{S'_A},r)$ as valid for the message~$\ket{M}$. If Bob wants to prove that Alice sent him this message, he will show $\ket{M}$ to Trent as input to the procedure \textsc{Proof-of-Origin} and send $\ket{S'_A}$ in step~2. Since $\ket{M}\neq\ket{M'}$, the arbitrator will agree with Alice, who will have succeeded in her disavowal.

For a message $\ket{M}$ to be repudiated, we consider $\ket{P}=E^{\text{KCCC}}_r\ket{M}$, and $\ket{P'}=X_1\ket{P}$ with $X_1$ being the gate $X$ applied on the first register. The state $\ket{M'}=(E^{\text{KCCC}}_r)^{-1}\ket{P'}$ is of the factorised form $\bigotimes_{i=1}^n\ket{M'_i}$ if $\ket{M}$ is. This observation follows from the commutation relations between the gates $\CNOT$ and~$X$:
\[
X_j\CNOT_{i,j}=\CNOT_{i,j}X_j \quad \text{and} \quad X_i \CNOT_{i,j} = \CNOT_{i,j} X_iX_j \quad \text{for } i\neq j.
\]
These relations imply that conjugating a product of Pauli matrices by $E^\CNOT_{r^1}$ yields, up to a global phase, another tensor product of Pauli matrices. The same observation also holds for the last two steps of encryption and therefore for $E^{\text{KCCC}}_r$.

Given~$\ket{M}$, Alice chooses $r$ so that $\ket{M'}$ is distinct from~$\ket{M}$. Writing the key $k_{BT}$ simply as $k = k^1\|k^2\|k^3$, the state sent back to Bob by the arbitrator is
\[
\ket{Y_{TB}} = E_{k}^{\text{KCCC}}\left(\ket{P'}\ket{S'_A}\right) = E^{\text{perm}}_{k^3}\circ H^{k^2}\circ E^{\CNOT}_{k^1}\left(\ket{P'}\ket{S'_A}\right).
\]
The first layer of encryption, $E^{\CNOT}_{k^1}$, depends on the permutation $k^1 = (k^1_{(1)}, \dots, k^1_{(2n)})$, with the image of $i$ written~$k^1_{(i)}$. We consider the event that $k^1_{(1)}=1$, that is, the permutation has $1$ as fixed point. If that event occurs, one has $X_1 E^{\text{\CNOT}}_{k^1} = E^{\text{\CNOT}}_{k^1} X_1$.

In the case where $E^{\text{transp}}$ is used (i.e., ``$\text{perm}$" is ``$\text{transp}$"), we consider the events that $k^2_1=0$ and $\tau_1\coloneq k^3_1\oplus k^3_{2n}=0$. These events are independent and each of them occurs with probability~$\frac{1}{2}$. In this case, Alice's strategy is to apply the operator $X_1$ to the state sent by Trent to Bob in step~6 of the verification procedure. If $k_{(1)}^1=1$, $k^2_1=0$ and $\tau_1=0$ occur, $E^{\text{transp}}_{k^3}$ and $H^{k^2}$ do not act on the first register and one has
\begin{align*}
X_1\ket{Y_{TB}} &= X_1 E_{k}^{\text{KCCC}}\left(\ket{P'}\ket{S'_A}\right)\\
&= X_1 E^{\text{transp}}_{k^3} H^{k^2} E^{\text{\CNOT}}_{k^1} \left(\ket{P'}\ket{S'_A}\right)\\
&= E^{\text{transp}}_{k^3} H^{k^2} E^{\text{\CNOT}}_{k^1} X_1 \left(\ket{P'}\ket{S'_A}\right)\\
&= E_{k}^{\text{KCCC}}\left(\ket{P}\ket{S'_A}\right).
\end{align*}
The state that Bob obtains by decryption is $\ket{P}\ket{S'_A}$. It will pass Bob checks (as he has both $\ket{P}$ and $\ket{R_{AB}}$) and Bob will accept the message $\ket{M}$ as legitimate. However, it can be later repudiated by Alice as Bob has $\ket{S'_A}$ and not~$\ket{S_A}$. The overall probability of such a successful repudiation is lower-bounded by
\[
\Pr\left(\text{repudiation}\right) \geqslant \Pr\left(k^1_{(1)}=1\right) \cdot \Pr\left(k^2_1 = 0\right) \cdot \Pr\left(\tau_1 = 0\right) = \frac{1}{8n}.
\]

In the case where $E^{\text{rot}}$ is used (i.e., ``$\text{perm}$" is ``$\text{rot}$"), we now consider the events $k^2_1=0$ and $\tau\coloneq (\sum_{i=1}^{2n} k^3_i)\pmod{2n} = n$. These events are independent of each other, and one has $\Pr(k^2_1=0)=\frac{1}{2}$ and $\Pr(\tau=n) = 2^{-2n}\cdot\binom{2n}{n} \geqslant \dfrac{1}{2\sqrt{n}}$ (using Lemma~7 of Chapter~10 of~\cite{MS81}). In this case, Alice's strategy is to apply the operator~$X_{n+1}$, defined as the gate $X$ applied to the $(n+1)^\text{th}$ register, to the state sent by Trent to Bob in step~6 of the verification procedure. If $k_{(1)}^1=1$, $k^2_1=0$ and $\tau=n$ occur, one has
\begin{align*}
X_{n+1}\ket{Y_{TB}} 
    &= X_{n+1} E_{k}^{\text{KCCC}}\left(\ket{P'}\ket{S'_A}\right)\\
    &=X_{n+1} E^{\text{rot}}_{k^3} H^{k^2} E^{\text{\CNOT}}_{k^1} \left(\ket{P'}\ket{S'_A}\right)\\
    &=E^{\text{rot}}_{k^3} X_1 H^{k^2} E^{\text{\CNOT}}_{k^1} \left(\ket{P'}\ket{S'_A}\right)\\
    &=E^{\text{rot}}_{k^3} H^{k^2} E^{\text{\CNOT}}_{k^1} X_1\left(\ket{P'}\ket{S'_A}\right)\\
    &= E_{k}^{\text{KCCC}}\left(\ket{P}\ket{S'_A}\right).
\end{align*}
In this case, similarly as above, Alice succeeds in repudiating her message. The success probability is lower-bounded by
\[
\Pr\left(\text{repudiation}\right) \geqslant \Pr\left(k_{(1)}^1=1\right) \cdot \Pr\left(k^2_1 = 0\right) \cdot \Pr\left(\tau=n\right) \geqslant \dfrac{1}{8n\sqrt{n}}.
\]

Although the authors of~\cite{ZSZJ17} did not bound the probability of a disavowal, the claim that it is ``\emph{obviously impossible}" and that Alice cannot disavow is unfounded. In any case, such a probability should decrease exponentially (or at least as a negligible function, with respect to some security parameter), which is not the case here.

Even though the attack uses the fact that the permutation $k^1$ has a fixed point, we stress that there exist similar attacks even if we restrict $k^1$ to permutations without any fixed points. In this article, we have chosen the simplest attack that highlights the flaws in the design.

\paragraph{False allegation by Alice.} As already mentioned in Subsection~\ref{subsec:scheme_CNOT_1:description}, Alice has a trivial false allegation strategy consisting in requiring a \textsc{Proof-of-Receipt} procedure for any message $\ket{\overline{M}}$ of her choice. As this procedure does not depend on the input, it succeeds with probability~$1$.

\paragraph{Forgery by Bob.} Suppose that Alice trustfully signs the message $\ket{M}$ by sending $E^{\text{KCCC}}_{k_{AB}}\big(\ket{P},\ket{R_{AB}},\ket{S_A}\big)$ to Bob. We denote the key $k_{AT}$ as $k=k^1\|k^2\|k^3$.

If one chooses the controlled pairwise transposition for the final step of the encryption (i.e., ``$\text{perm}$" is ``$\text{transp}$"), we consider the three independent events $k^1_{(1)}=1$ (i.e., $k^1$ fixes~$1$), $k^2_1=0$ and $\tau_1 \coloneq k^3_1 \oplus k^3_n=0$ which occur respectively with probability $\frac{1}{n}$, $\frac{1}{2}$ and~$\frac{1}{2}$ (or $1$ if $n=1$). In that case, Bob's strategy consists in applying $X_1$ to both $\ket{P}$ and $\ket{S_A}$ in the beginning of the verification procedure to obtain $\ket{P'}$ and~$\ket{S_A'}$. If $k^1_{(1)}=1$, $k^2_1=0$ and $\tau_1=0$ hold, one has
\begin{align*}
\ket{S_A'} &= X_1\ket{S_A}\\
&= X_1 E^{\text{KCCC}}_k\ket{P}\\
&= X_1 E^{\text{transp}}_{k^3} H^{k^2} E^{\text{CNOT}}_{k^1}\ket{P}\\
&= E^{\text{transp}}_{k^3} H^{k^2} E^{\text{CNOT}}_{k^1}X_1\ket{P}\\
&= E^{\text{KCCC}}_k\ket{P'}.
\end{align*}
Therefore, in that case, if Bob sends $\ket{Y_{BT}'}=E^{\text{KCCC}}_{k_{BT}}\big(\ket{P'},\ket{S_A'}\big)$ to Trent, the later will write $v_T=1$ on the public board. Then, Bob can write $v_B=1$ at step~8 of the \textsc{Verify} procedure. At the end, he will save $(\ket{M'},\ket{S_A'},r)$ where $\ket{M'}=\left(E^{\text{KCCC}}_r\right)^{-1}\ket{P'}$. During a \textsc{Proof-of-Origin} procedure, he will thus be able to prove that Alice sent him the message~$\ket{M'}$, which is not true if $\ket{M}\neq\ket{M'}$, that is if $\ket{P}\neq\ket{P'}$ (up to a global phase). Since $\ket{P'}=X_1\ket{P}$, this last condition occurs exactly when $\ket{P}$ is neither of the form $\ket{+}\ket{Q}$ nor of the form $\ket{-}\ket{Q}$. Fixing all bits of $r=r^1\|r^2\|r^3$ except that of $r^2$ influencing the first qubit of~$\ket{P}$, this happens for at least one of the two choices for that bit, leading to $\Pr(\ket{P}\neq\ket{P'})\geqslant\frac{1}{2}$. The overall success probability of this forgery attack is lower-bounded by
\[
\Pr\left(\text{forgery}\right) \geqslant \Pr\left(k^1_{(1)}=1\right) \cdot \Pr\left(k^2_1=0\right) \cdot \Pr\left(\tau_1=0\right) \cdot \Pr\left(\ket{P}\neq\ket{P'}\right) \geqslant \frac{1}{8n}
\]
which is not negligible.

If one chooses the rotation as controlled permutation (i.e., ``$\text{perm}$" is ``$\text{rot}$"), we consider the three independent events $k^1_{(1)}=1$, $k^2_1=0$ and $\tau \coloneq (\sum_{i=1}^{n} k^3_i)\pmod{n} = \lfloor \frac{n}{2}\rfloor$. Using Lemma~7 of Chapter~10 of~\cite{MS81}, the third event occurs with probability $\Pr(\tau=\lfloor\frac{n}{2}\rfloor) = 2^{-n}\cdot\binom{n}{\lfloor\frac{n}{2}\rfloor} \geqslant \dfrac{1}{2\sqrt{n}}$ if $n>1$ (or probability $1$ if $n=1$). Bob can use a strategy similar to that above, applying the gate $X$ to the first qubit of $\ket{P}$ and, if $n>1$, to the $(n-\lfloor\frac{n}{2}\rfloor+1)^\text{th}$ qubit of $\ket{S_A}$ (or its first qubit if $n=1$) to get $\ket{P'}$ and $\ket{S_A'}$ respectively. The overall success probability of this forgery attack is lower-bounded by
\[
\Pr\left(\text{forgery}\right) \geqslant \Pr\left(k^1_{(1)}=1\right) \cdot \Pr\left(k^2_1=0\right) \cdot \Pr\left(\tau=\Bigl\lfloor\frac{n}{2}\Bigr\rfloor\right) \cdot \Pr\left(\ket{P}\neq\ket{P'}\right) \geqslant \frac{1}{8n\sqrt{n}}
\]
which is again not negligible.

For the same reason given in Alice's repudiation, it is not hard to see that, in both cases, the message $\ket{M'}$ that Bob pretends Alice sent is of the factorised form $\bigotimes_{i=1}^n\ket{M'_i}$ if $\ket{M}$ is.

\FloatBarrier
%

\section{Cryptanalysis of a scheme based on GHZ states}
\label{sec:GHZ_3}

In this section, we analyse the scheme proposed in~\cite{DXYS22}. Its security is based on GHZ-like states shared between parties, but which are not used in a teleportation protocol. Instead, the parties verify that the three registers of the GHZ states have a predetermined XOR-sum in the computational basis. The scheme builds upon the protocol of~\cite{ZK20}, which is shown by the authors of~\cite{DXYS22} to be trivially insecure: Bob can apply $X$ gates to the message and its signature to forge a valid signed message. The improved protocol aims to address this insecurity.

\subsection{Description of the protocol}

\begin{algorithm}[p!]
\begin{algorithmic}
\Procedure{Init}{}
\State 1. $\T$: prepare and distribute $n$ GHZ states $H^{\otimes 3}\ket{GHZ}_{ABT}$.
\State 2. $\A\leftrightarrow\T$: $k_A\in\{0,1\}^n$.
\EndProcedure
\vspace{1mm}

\Procedure{Sign }{\text{on a classical message} $m\in\{0,1\}^n$}
\State 1. $\A$: $t=G_1(m\|k_A)$ and $s=G_2(m\|k_A)$.
\State 2. $\A$: $\ket{P^0_A}=\ket{P^1_A}=H^{t}X^{s}\ket{m}$.
\State 3. $\A$: $\CNOT (\ket{A}\ket{P^i_A})=\ket{A}\ket{A\oplus P^i_A}$ for each $i\in\{0,1\}$.
\State 4. $\A$: $\left(\ket{A\oplus P^0_A},\ket{A\oplus P^1_A},m\right)\to\B$ \hfill // secured using decoy states.
\State 5. $\B$: store $\left(\ket{B},\ket{A\oplus P^0_A},m\right)$.
\EndProcedure
\vspace{1mm}

\Procedure{Verify}{}
\State 1. $\B$: $\CNOT(\ket{B}\ket{A\oplus P^1_A})=\ket{B}\ket{B\oplus A\oplus P^1_A}$.
\State 2. $\B$: $\left(\ket{B\oplus A\oplus P^1_A},\ket{m}\right) \to \T$ \hfill // secured using decoy states.
\State 3. $\T$: $\CNOT\left(\ket{T}\ket{B\oplus A\oplus P^1_A}\right) = \ket{T}\ket{T\oplus B\oplus A\oplus P^1_A}\eqcolon \ket{T}\ket{P'_A}$.
\State 4. $\T$: $m_T =$ measure of $\ket{m}$ along $Z$ basis.
\State 5. $\T$: $t'=G_1(m_T\|k_A)$ and $s'=G_2(m_T\|k_A)$.
\State 6. $\T$: $m' =$ measure of $X^{s'}H^{t'}\ket{P'_A}$ along $Z$ basis.
\State 7. $\T$: $m'=m_T$ ? store $\left(\ket{T},\ket{P'_A},m_T\right)$ and \textbf{accept} : store $(m_T=\bot)$ and \textbf{reject}.
\EndProcedure

\Procedure{Proof-of-Origin }{\text{on input} $\overline{m}\in\{0,1\}^n$}
\State 1. $\T$: $m_T=\bot$ ? \textbf{favour~$\A$} : continue. 
\State 2. $\B$: $\CNOT(\ket{B}\ket{A\oplus P^0_A})=\ket{B}\ket{B\oplus A\oplus P^0_A}$.
\State 3. $\B$: $\ket{B\oplus A\oplus P^0_A} \to \T$.
\State 4. $\T$: $\CNOT\left(\ket{T}\ket{B\oplus A\oplus P^0_A}\right) = \ket{T}\ket{T\oplus B\oplus A\oplus P^0_A}\eqcolon \ket{T}\ket{P''_A}$.
\State 5. $\T$: $\overline{t}=G_1(\overline{m}\|k_A)$ and $\overline{s}=G_2(\overline{m}\|k_A)$.
\State 6. $\T$: $m'' =$ measure of $X^{\bar{s}}H^{\bar{t}}\ket{P''_A}$ along $Z$ basis.
\State 7. $\T$: $m''=\overline{m}$ ? \textbf{favour~$\B$} : \textbf{favour~$\A$}.
\EndProcedure

\Procedure{Proof-of-Receipt }{\text{on input} $\overline{m}\in\{0,1\}^n$}
\State 1. $\T$: $m_T=\bot$ ? \textbf{favour~$\B$} : continue. 
\State 2. $\T$: $\overline{t}=G_1(\overline{m}\|k_A)$ and $\overline{s}=G_2(\overline{m}\|k_A)$.
\State 3. $\T$: $m' =$ measure of $X^{\overline{s}}H^{\overline{t}}\ket{P'_A}$ along $Z$ basis.
\State 4. $\T$: $m'=\overline{m}$ ? \textbf{favour~$\A$} : \textbf{favour~$\B$}.
\EndProcedure
\end{algorithmic}
\caption[AQS Protocol~\ref{alg:AlgoDing}, of Ding, Xin, Yang and Sang]{AQS protocol of~\cite{DXYS22} based on GHZ states.}
\label{alg:AlgoDing}
\end{algorithm}

The AQS protocol of~\cite{DXYS22}, described in Protocol~\ref{alg:AlgoDing}, is designed to sign classical messages.

The protocol uses GHZ-like states that are shared between parties during initialisation. Specifically, given a GHZ state $\ket{GHZ}=\frac{1}{\sqrt{2}}(\ket{000}+\ket{111})$, the state
\[
H^{\otimes 3}\ket{GHZ} = \frac{1}{\sqrt{2}}(\ket{+++}+\ket{---}) = \frac{1}{2}(\ket{000}+\ket{011}+\ket{101}+\ket{110})
\]
is created and shared by sending one of the three registers to each involved party. To sign a classical message of length~$n$, $n$ such states are produced and distributed. We denote the $n$ qubit state received by Alice (respectively by Bob, by Trent) as~$\ket{A}$ (respectively $\ket{B}$,~$\ket{T}$). This step is assumed to be done securely by the arbitrator. The scheme relies on two one-way hash functions, $G_1,G_2\colon\{0,1\}^\ast\to \{0,1\}^n$, which are considered secure.

By abuse of notation, for two $n$ qubit states $\ket{P}$ and~$\ket{Q}$, we write $\CNOT(\ket{P},\ket{Q})$ simply by $\ket{P}\ket{P\oplus Q}$ where $\CNOT$ is defined for $P_1,\dots,P_n,Q_1,\dots,Q_n\in\{0,1\}$ by
\[
\CNOT\left(\bigotimes_{i=1}^n\ket{P_i},\bigotimes_{i=1}^n\ket{Q_i}\right)=\left(\bigotimes_{i=1}^n\ket{P_i}\right) \otimes\left(\bigotimes_{i=1}^n\ket{P_i\oplus Q_i}\right)
\]
and extended linearly. We stress that, following~\cite{DXYS22}, we use throughout this section a notation in which possibly entangled states are written as factorised kets. Thus, expressions such as $\ket{A}\ket{B}\ket{T}$ or $\ket{P}\ket{P\oplus Q}$ should not be interpreted as tensor products of independent pure states, but as a convenient description of the relevant registers of a possibly entangled state. As an example, during step~3 of the \textsc{Verify} procedure, the state $\CNOT(\ket{T}\ket{B\oplus A\oplus P_A^1})$ may be entangled, but we still denote it by $\ket{T}\ket{P'_A}$ and refer to operations on the second register as acting on~$\ket{P'_A}$.

\paragraph{Ambiguities.} The proposed protocol contains several ambiguities. We list them below and state the conventions we adopt to resolve them.

In step~4 of the \textsc{Sign} procedure and in step~2 of the \textsc{Verify} procedure, quantum messages are sent, and these communications are secured ``by the eavesdropping detection technology~\cite{LLDZLZ07}", which is based on decoy states. The authors do not make precise statement on this matter (type of decoys, number...). As our attacks do not rely on that, we do not make it more explicit in Protocol~\ref{alg:AlgoDing}. It is implicit that the protocol is aborted and the signature rejected if the decoy checks fail. Additionally, we note that the message $m$ is sent classically to Bob during step~4 of the \textsc{Sign} procedure, and then as a quantum state during the \textsc{Verify} one. Besides the fact that we do not see the added value of this discrepancy, not protecting the integrity of $m$ in the \textsc{Sign} procedure enlarges the attack surface. 

Regarding the registers and messages that are stored by the parties, we assume that both Bob and the arbitrator keep their GHZ register: Bob stores $(\ket{B}, \ket{A\oplus P^0_A},m)$ and Trent keeps $(\ket{T}\ket{P'_A},m_T)$ in case he accepts. We note that Trent must reconstruct the state $\ket{P'_A}$ as $H^{t'}X^{s'}\ket{m'}$ after its measurement during the step~6 of the \textsc{Verify} procedure. The protocol does not describe what happens if he rejects the signature (at the end of the \textsc{Verify} procedure or during decoy verification). For the sake of simplicity, we assume he stores a rejection token $\bot$ used in the later procedures.

The paper~\cite{DXYS22} does not specify the procedures \textsc{proof-of-Origin} and \textsc{Proof-of-Receipt}. Several descriptions are possible here, the simplest being that, since Trent saved $m_T$ at the end of a successful verification procedure, he checks whether $m_T$ is equal to the input $\overline{m}$ of the procedure \textsc{proof-of-Origin} or \textsc{Proof-of-Receipt}, and gives a judgement in favour of the complainant if and only if $m_T=\overline{m}$. However, this would make useless the storage of quantum states by Bob and Trent at the end of the \textsc{Sign} and \textsc{Verify} procedures, respectively. Based on the statements in~\cite{DXYS22} that ``Bob stores $\{\ket{A\oplus P_A},m\}$ as Alice's signature", ``If Bob refuses that he has ever received the quantum signature, Trent can judge the fact by checking the sequence $\{\ket{P'_A},m\}$" and ``Trent can decode $\ket{P'_A}$ [...] by performing the signature verification", we opt for more complex procedures \textsc{proof-of-Origin} and \textsc{Proof-of-Receipt} in Protocol~\ref{alg:AlgoDing} that do not compare $m_T$ with~$\overline{m}$ (hence that do not require Trent to keep a transcript of the messages that have been signed).

\subsection{Analysis of the security claims}
\label{subsec:analyse_2}

The authors of~\cite{DXYS22} claim that their scheme offers resistance to forgery and non-repudiation. While we have found no attack regarding the forgery, we note that the scheme is probably not information-theoretically secure (ITS) as its security relies on the security of hash functions. As there exist classical schemes to sign classical messages with computational security, this scheme may not be the most efficient solution.

In the following, we focus on security claims regarding disavowal by the sender or the receiver, as well as a false allegation strategy for Alice if equipped with unbounded resources.

\paragraph{Repudiation by Alice.} In step~4 of the \textsc{Sign} procedure, suppose that Alice sends $\left(\ket{\text{gibberish}},\ket{A\oplus P^1_A},m\right)$. In that case, the message will first be deemed valid by the arbitrator during the \textsc{Verify} procedure. If later Alice repudiates her message and that Bob confronts her with the signature during a \textsc{Proof-of-Origin} procedure on input~$m$, the arbitrator will (most probably) reject the signature Bob kept since $\ket{P''_A}$ is not related to $m$ anymore. Therefore, Alice will have successfully disavowed. However, note that this attack does not work if the simpler \textsc{Proof-of-Origin} procedure consisting in the single step ``$\T$: $\overline{m}=m_T$ ? \textbf{favour~$\B$} : \textbf{favour~$\A$}" is chosen instead, but in which case the quantum information stored by Bob is unused.

\paragraph{False allegation by Alice with unbounded resources.} 
We consider the case of a malicious Alice with unbounded resources. She wishes to find two distinct messages $m,\overline{m}$ yielding the same~$\ket{P^0_A}$. For convenience, we introduce the function $G'_2(m,k) = m\oplus G_2(m\|k)$, and recall that $\ket{P^0_A}$ is given by 
\[\ket{P^0_A} = H^{G_1(m\|k_A)} X^{G_2(m\|k_A)}\ket{m} = H^{G_1(m\|k_A)} X^{G'_2(m,k_A)}\ket{0}.\]
To succeed in her false allegation attempt, Alice uses her unbounded resources to find two distinct messages $m,\overline{m}$ such that $G_1(m\|k_A)=G_1(\overline{m}\|k_A)$ and $G_2'(m,k_A)=G_2'(\overline{m},k_A)$. Alice can then claim she has sent $\overline{m}$ while she sent~$m$. For random functions $G_1$ and~$G_2'$, the probability that there exists such messages is non-negligible. To evaluate this, we consider the function
$$F_{k_A}\colon\{0,1\}^n\to \{0,1\}^{2n},\quad m\mapsto (G_1(m\|k_A), G_2'(m,k_A)),$$
and we apply the standard birthday argument. The probability that we are evaluating is the collision probability of~$F_{k_A}$:
$$\Pr[\exists \ m,\overline{m}| F_{k_A}(m)=F_{k_A}(\overline{m})] = 1-\prod_{i=0}^{2^{n}-1}\left(1-\frac{i}{2^{2n}}\right)\geqslant 1 -e^{-\frac{2^n(2^n-1)}{2^{2n+1}}}.$$
This probability tends toward $1-e^{-1/2}\simeq 39\%$ for large~$n$. We note that if Alice has some control on~$k_A$ (which is usually the case in standard QKD protocols such as BB84), then she can find such a collision with overwhelming probability.

As for the above repudiation attack, we note that this attack does not work if the simpler \textsc{Proof-of-Receipt} procedure consisting in the single step ``$\T$: $\overline{m}=m_T$ ? \textbf{favour~$\A$} : \textbf{favour~$\B$}" is chosen instead.

\paragraph{Disavowal by Bob.} During the course of the protocol, the receiver has access to the complete cleartext message before the arbitrator performs his part of the verification. Therefore, if the message is not to his liking, Bob can decide not to forward the message and the quantum states to Trent. This simple disavowal strategy is not considered by the authors as the protocol has not been terminated. However, this is a winning strategy that we consider valid. Note that in the most recent AQS schemes, this issue is resolved by delaying the full disclosure of the message to the recipient at the last step of the verification (see e.g., \cite{ZQ10, LS15,ZK20} or Protocol~\ref{alg:Algo_CNOT}).

\FloatBarrier
%

\section{Cryptanalysis of a scheme based on controlled teleportation}
\label{sec:scheme_controlled_teleportation_2}

In this section, we review the AQS protocol of Lu et al.~\cite{LLYH22} described in Protocol~\ref{alg:AlgoLu}, which is based on a five-qubit entangled system, and we discuss its security claims.

\subsection{Description of the protocol}

\begin{algorithm}[p!]
\begin{algorithmic}
\Procedure{Init}{}
\State 1. $\T$: $k_A \to \A$.
\State 2. $\T$: $k_B \to \B$.
\EndProcedure{}
\vspace{1mm}

\Procedure{Sign }{\text{on 4 copies $\ket{M^0}=\ket{M^1}=\ket{M^2}=\ket{M^3}$ of} $\ket{M}=\bigotimes_{i=1}^n\ket{M_i}$}
\State 1. $\A$: $\ket{R_A} = R_{k_A}\ket{M^0}$.
\State 2. $\A$: $\ket{\Omega}=\bigotimes_{i=1}^n \ket{\Omega_i}$ with $\ket{\Omega_i} = \ket{M^1_i}\otimes\ket{\xi}$.
\State 3. $\A$: share $\ket{\Omega}_{ATAATB} = \bigotimes_{i=1}^n \ket{\Omega_i}_{ATAATB}$ into $\ket{\Omega_A},\ket{\Omega_B}$ and~$\ket{\Omega_T}$. 
\State 4. $\A$: $\ket{\Omega_T} \to \T$, $\ket{\Omega_B} \to \B$ \hfill // secured using decoy states.
\State 5. $\A$: $\ket{\omega_A} = \ket{\Omega_A}$ after measurement according to the basis $\{\ket{\chi_\beta}\}_{\beta=0}^7$.
\State 6. $\A$: $\ket{S} = E_{k_A}\left(\ket{R_A},\ket{\omega_A}\right)$.
\State 7. $\A$: $(\ket{M^2}, \ket{M^3},\ket{S}) \to \B$.
\EndProcedure
\vspace{1mm}

\Procedure{Verify}{}
\State 1. $\B$: $\ket{Y_{BT}} = E_{k_B} (\ket{S},\ket{M^2}) \to \T$ \hfill // without decoy.
\State 2. $\T$: recover $\ket{S},\ket{M^2}$; from~$\ket{S}$, recover $\ket{R_A},\ket{\omega_A}$.
\State 3. $\T$: $\ket{\omega_T} = \ket{\Omega_T}$ after measurement according to Bell states basis $\{\ket{\Psi_\alpha}\}_{\alpha=0}^3$.
\State 4. $\T$: $\ket{R_A} = R_{k_A}\ket{M^2}\ ? \ \theta=1 : \theta=0$ \hfill // $\T$ has now 2 copies of~$\ket{R_A}$.
\State 5. $\T$: $\ket{S^0} = \ket{S^1} = E_{k_A}(\ket{R_A},\ket{\omega_A})$ \hfill // $\T$ uses his 2 copies of~$\ket{R_A}$.
\State 6. $\T$: choose a hash function $\mathrm{H}$ and calculate $\mathrm{H}(\ket{S^1})$.
\State 7. $\T$: $\ket{Y_{TB}} = E_{k_B} (\ket{S^0},\mathrm{H}(\ket{S^1}),\ket{\omega_A},\ket{\omega_T},\ket{\theta}) \to \B$ \hfill // without decoy.
\State 8. $\T$: store $(\omega_A,\theta)$.
\State 9. $\B$: recover $\ket{S^0}$, $\mathrm{H}(\ket{S^1})$, $\ket{\omega_A}$, $\ket{\omega_T}$, $\theta$; $\theta=0$ ? \textbf{reject} : continue.
\State 10. $\B$: recover $\ket{M^1}$ from $\ket{\Omega_B}$ using $\ket{\omega_A}$ and~$\ket{\omega_T}$.
\State 11. $\B$: $\ket{M^1} = \ket{M^3}$ ? continue : \textbf{reject}.
\State 12. $\B$: $\mathrm{H}(\ket{S^0}) = \mathrm{H}(\ket{S^1})$ ? store $(\ket{M^1},\ket{S^0},\ket{S^1})$ and \textbf{accept} : \textbf{reject}.
\EndProcedure

\Procedure{Proof-of-Origin }{\text{on input} $\ket{\overline{M}}=\bigotimes_{i=1}^n\ket{\overline{M_i}}$}
\State 1. $\T$: $\theta=1$ ? continue : \textbf{favour~$\A$}.
\State 2. $\B$: $\ket{S^0} \to \T$.
\State 3. $\T$: $\ket{\overline{R_A}} = R_{k_A}\ket{\overline{M}}$.
\State 4. $\T$: $E_{k_A}\left(\ket{\overline{R_A}},\ket{\omega_A}\right) = \ket{S^0}$ ? \textbf{favour~$\B$} : \textbf{favour~$\A$}.
\EndProcedure

\Procedure{Proof-of-Receipt }{\text{on input} $\ket{\overline{M}}=\bigotimes_{i=1}^n\ket{\overline{M_i}}$}
\State 1. $\T$: $\theta=1$ ? continue : \textbf{favour~$\B$}.
\State 2. $\B$: $\ket{S^1} \to \T$.
\State 3. $\T$: $\ket{\overline{R_A}} = R_{k_A}\ket{\overline{M}}$.
\State 4. $\T$: $E_{k_A}\left(\ket{\overline{R_A}},\ket{\omega_A}\right) = \ket{S^1}$ ? \textbf{favour~$\A$} : \textbf{favour~$\B$}.
\EndProcedure
\end{algorithmic}

\caption[AQS Protocol~\ref{alg:AlgoLu}, of Lu, Li, Yu and Han]{AQS protocol of~\cite{LLYH22}, based on controlled teleportation.}
\label{alg:AlgoLu}
\end{algorithm}

\paragraph{Two encryptions.} This scheme involves two encryption methods. The first one is the standard quantum one-time pad~\cite{BR03} defined as
\[
E_k\ket{M} = \bigotimes_{i=1}^n X^{k_{2i-1}}Z^{k_{2i}}\ket{M_i}
\]
for a key $k\in\{0,1\}^{2n}$ and an $n$-qubit state $\ket{M}=\bigotimes_{i=1}^n \ket{M_i}$ (and extended linearly). The other encryption method used in Protocol~\ref{alg:AlgoLu} is that of~\cite{ZQ10}, defined as
\[
R_k\ket{M} = \bigotimes_{i=1}^n X^{k_i}Z^{k_i \oplus 1}\ket{M_i}
\]
for a key $k\in\{0,1\}^n$ and an $n$-qubit state $\ket{M}=\bigotimes_{i=1}^n \ket{M_i}$ (and extended linearly).

\paragraph{Two bases and a five-qubit state.} The Bell state basis is denoted by $\{\ket{\Psi_\alpha}\}_{\alpha=0}^3$ and defined as
\[
\ket{\Psi_0}=\frac{1}{\sqrt{2}}\left(\ket{00}+\ket{11}\right),\quad \ket{\Psi_\alpha}= \left(Z^{b_1}\otimes X^{b_2}\right)\ket{\Psi_0}, 
\]
where $b_1\|b_2$ is the binary representation of~$\alpha$. One also needs the von Neumann basis $\{\ket{\chi_\beta}\}_{\beta=0}^7$ defined by
\[
\ket{\chi_0}=\frac{1}{\sqrt{2}}\left(\ket{000}+\ket{111}\right),\quad \ket{\chi_\beta}= \left(X^{b_1}\otimes X^{b_2} \otimes Z^{b_3}\right)\ket{\chi_0}, 
\]
where $b_1\|b_2\|b_3$ is the binary representation of~$\beta$ (the states $(\chi_0,\chi_1,\chi_2,\chi_3,\chi_4,\chi_5,\chi_6,\chi_7)$ are denoted in~\cite{LLYH22} as $(\chi^1,\chi^2,\chi^5,\chi^6,\chi^7,\chi^8,\chi^3,-\chi^4)$).

The five-qubit state $\ket{\xi}$ used in the protocol is
\[
\ket{\xi} = 
\frac12\left(\ket{100}\ket{\Psi_0}+\ket{111}\ket{\Psi_1}+\ket{001}\ket{\Psi_2}+\ket{010}\ket{\Psi_3}\right).
\]
This state is used in a teleportation-like protocol of a qubit~$\ket{\varphi}$. First the joint state $\ket{\Omega}= \ket{\varphi}\otimes \ket{\xi}$ is created. It is composed of six entangled states which are shared between the parties. We note $\ket{\Omega}_{ATAATB}$ to indicate this sharing. This notation means that Alice has the first, the third and the fourth registers composing~$\ket{\Omega}$, and similarly for Trent and Bob. The resulting states are noted $\ket{\Omega_A},\ket{\Omega_T}$ and~$\ket{\Omega_B}$. Second, Alice and the arbitrator both perform a measurement of their state and note the respective collapsed states $\ket{\omega_A}$ and~$\ket{\omega_T}$.
Finally, after the measurement of the first five registers, Bob can recover the initial message $\ket{\varphi}$ by applying the correct transformations (depending on $\ket{\omega_A}$ and~$\ket{\omega_T}$) on the sixth register that he owns,~$\ket{\Omega_B}$.

As in the previous protocol, some quantum communications are secured using decoy states. If the decoy checks fail upon reception of a quantum message, the signature is rejected and Trent stores $\theta=0$ before aborting. As our attack does not exploit a weakness related to these decoys, we only briefly mention them in the protocol description. In particular, we do not describe the exchange of polynomials in the \textsc{Init} procedure aimed at selecting the state of these decoys. Note that only the states that share the teleportation state $\ket{\Omega}$ are protected with decoys, and not all quantum messages. 

\paragraph{Ambiguities.} We note that there are multiple ambiguities in~\cite{LLYH22}. The first one concerns the number of copies of the initial message required. The article mentions that three copies are needed. However, as can be seen from Protocol~\ref{alg:AlgoLu}, one copy of the message is used to construct~$\ket{R_A}$, one copy is entangled with the five-qubit states, and two are sent to Bob at the end of the signing procedure. Moreover, the protocol does not take into account that multiple copies should be used for the equality tests in steps~4, 11 and~12 of the \textsc{Verify} procedure.

A second ambiguity concerns the hash function $\mathrm{H}$ and its usage. The protocol requires the arbitrator to use a hash function on the signature~$\ket{S^1}$. This hash is claimed to ensure the integrity of the signature. However, the nature of $\mathrm{H}$ is not specified in~\cite{LLYH22}. We consider two cases: either $\mathrm{H}(\ket{S})$ is a classical bit string computed from~$\ket{S}$, or $\mathrm{H}$ is a quantum unitary operation applied on $\ket{S}$ such that $\ket{h} = \mathrm{H}(\ket{S})$ is a quantum state. In the first case, Bob would be unable to send $\ket{S^0}$ to Trent in a \textsc{proof-of-Origin} procedure. Another problem is that such a classical string should be obtained after a measurement of~$\ket{S}$. As this state is unknown to Bob, the result of $\mathrm{H}$ would be random, and likely leading to a rejection in step~12 of the \textsc{Sign} procedure. For these reasons, we reject this first option and consider the second one, where $\mathrm{H}$ is a quantum unitary operation. Note that in that case, Trent does not hold a copy of $\mathrm{H}(\ket{S^1})$ after the \textsc{Verify} procedure, which causes issues for Trent to judge whether Bob has indeed received the message as explained below. Furthermore, since $\mathrm{H}$ is unitary, it is easily invertible, which is anyway mandatory for Bob to re-obtain $\ket{S^{0}}$ as a signature at the last step of the \textsc{Verify} procedure. In addition, the scheme specifies that the arbitrator unilaterally chooses the hash function. We assume that this choice is publicly announced to Bob after the reception of~$\ket{Y_{TB}}$. The length of the hash output is not given. As a unitary, it cannot be smaller than the one of the input and we therefore suppose it is length-preserving.

It is unclear whether classical information (e.g., the bit~$\theta$) is sent classically or as a quantum state~($\ket{\theta}$). We opt for the second option, as the authors use the quantum one-time pad which is defined for quantum states, and assume that the recipient implicitly measures the quantum state in the appropriate basis to recover classical information. Note that this choice does not influence our attacks.

A fourth ambiguity concerns the keys $k_A$ and~$k_B$. In the description of the scheme in~\cite{LLYH22}, their length is not given explicitly. In Appendix~A of that paper, it appears in an example that $k_A,k_B\in\{0,1\}^{2n}$. However, in step~6 of the \textsc{Sign} procedure, Alice computes $E_{k_A}\left(\ket{R_A},\ket{\omega_A}\right)$ where $\ket{R_A}$ is $n$-qubit long and $\ket{\omega_A}$ is $3n$-qubit long. This requires $k_A$ to be at least $8n$-bit long. Moreover, in step~7 of the \textsc{Verify} procedure, Trent computes $E_{k_B} (\ket{S^0},\mathrm{H}(\ket{S^1}),\ket{\omega_A},\ket{\omega_T},\ket{\theta})$ where $\ket{S^0}$ and $\mathrm{H}(\ket{S^1})$ are both $4n$-qubit long, $\ket{\omega_A}$ and $\ket{\omega_T}$ have respective length $3n$ and $2n$ qubits, and $\theta$ is a single bit encoded in a qubit. One thus needs that $k_B$ is at least $(26n+2)$-bit long. In addition, the same key $k_A$ is used for two different encryptions, $\ket{R_A}$ and~$\ket{S}$ and, similarly, $k_B$ is used in $\ket{Y_{BT}}$ and $\ket{Y_{TB}}$. All these encryptions require a different key length (i.e., $n$ and $8n$ for~$k_A$, and $10n$ and $26n+2$ bits for~$k_B$). In order to use independent keys and increase security, we assume that one chooses $k_A\in\{0,1\}^{9n}$ and $k_B\in\{0,1\}^{36n+2}$, where their first part is used for the first encryption and the last part for the second one.

The authors of~\cite{LLYH22} do not describe the procedures \textsc{Proof-of-Origin} and \textsc{Proof-of-Receipt}. We make a proposition for both of them in Protocol~\ref{alg:AlgoLu}. This requires that Trent saves~$\theta$, $k_A$ and the measurement results~$\ket{\omega_A}$, and that Bob keeps $\ket{S^0}$, $\ket{S^1}$ and~$\ket{M}$. Our proposition for the \textsc{Proof-of-Origin} procedure is quite intuitive, but we did not find a good solution for the \textsc{Proof-of-Receipt} procedure. Indeed, similarly to Protocol~\ref{alg:Algo_CNOT}, we could have required Trent to check only $\theta=1$, but this would have led to a trivial false allegation attack by Alice who could have pretended that Bob received any message since this procedure does not depend on~$\ket{\overline{M}}$. We rather opt for another solution, which is similar to the \textsc{Proof-of-Origin} procedure, but which is trivially deniable by Bob as he could send a garbage state in step~2.

Another issue concerns the space of messages that Alice can sign. Similarly to Protocol~\ref{alg:Algo_CNOT}, Alice cannot sign all $n$-qubit messages, but only those of the factorised form $\ket{M}=\bigotimes_{i=1}^n\ket{M_i}$. This restriction is due to the quantum teleportation which is realised qubit-wise. However, it is not clear to us if the protocol is claimed to be protected against adversaries that do not follow this requirement, for example, if Alice starts the \textsc{Sign} procedure with an entangled message. Although we think these are legitimate attacks, we make sure to use factorised messages in our attacks.

\subsection{Analysis of the security claims}
\label{subsec:analyse_3}

The authors of~\cite{LLYH22} claim that their protocol achieves unforgeability and non-repudiation for both the sender and the receiver. In this section, we show that these claims are not valid.

\paragraph{Repudiation and false allegation by Alice.} In the beginning of the \textsc{Sign} procedure, Alice chooses $\ket{M^0}=\ket{M^2}\neq\ket{M^1}=\ket{M^3}$. In that case, Bob gets two different messages $\ket{M^2}$ and $\ket{M^3}$ at the end of the \textsc{Sign} procedure, but all checks succeed in the \textsc{Verify} procedure. Since $\ket{S^0}$ is computed from~$\ket{M^2}$, Bob will be able to prove in a \textsc{Proof-of-Origin} procedure that he received it, but not that he received~$\ket{M^3}$, which is the copy of the message that he could used. This strategy also allows Alice to prove that Bob received $\ket{M^2}$ in a \textsc{Proof-of-Receipt} procedure, although he believes he received~$\ket{M^3}$.

\paragraph{Disavowal by Bob.} Bob has two trivial disavowal strategies. The first is similar to that of Protocol~\ref{alg:AlgoDing}. As Bob receives the full message in the last step of the \textsc{Sign} procedure, he can decide to stop the protocol. The solution to this issue is to design a protocol for which the delivery of the intended message is delayed until the end of the protocol, as in~\cite{ZQ10}.

For the second one, Bob simply sends another state than $\ket{S^1}$ in step~2 of the \textsc{Proof-of-Receipt} procedure.

\paragraph{Forgery by Bob.} We show here a forgery attack that Bob can apply to any qubit of the message~$\ket{M}$. Let us suppose that Bob attacks the first qubit. At the end of the \textsc{Sign} procedure, Bob receives from Alice two copies $\ket{M^2}=\ket{M^3}$ of~$\ket{M}$, as well as $\ket{S}=E_{k_A}(\ket{R_A},\ket{\omega_A})$ where $\ket{R_A}=R_{k_A}\ket{M^0}$. Let $a$ be the first bit of the part of $k_A$ used to compute~$\ket{R_A}$, and $b$ and $c$ be the first and second bits of the part of $k_A$ used to compute~$\ket{S}$. In view of our assumptions we made to resolve the ambiguities, one has $a=(k_A)_1$, $b=(k_A)_{n+1}$ and $c=(k_A)_{n+2}$. The first qubit of $\ket{R_A}$ is $X^aZ^{a\oplus 1}\ket{M^0_1}$, and the first qubit of $\ket{S}$ is $X^bZ^cX^aZ^{a\oplus 1}\ket{M^0_1}$.

We suppose that, at the start of the \textsc{Verify} procedure, Bob applies an $X$ gate to the first qubit of $\ket{M^2}$ and to the first qubit of~$\ket{S}$. They thus become $X\ket{M^2_1}$ and $X^{b\oplus 1}Z^cX^aZ^{a\oplus 1}\ket{M^0_1}$, respectively, while the other qubits are left unchanged. When Trent decrypts $\ket{S}$ to get~$\ket{R_A}$, he obtains the correct $\ket{R_A}$ except for the first qubit which is now
\begin{align*}
Z^cX^bX^{b\oplus 1}Z^cX^aZ^{a\oplus 1}\ket{M^0_1} &= Z^cXZ^cX^aZ^{a\oplus 1}\ket{M^0_1}\\
&= (-1)^cXZ^cZ^cX^aZ^{a\oplus 1}\ket{M^0_1}\\
&= (-1)^cX^{a \oplus 1}Z^{a\oplus 1}\ket{M^0_1}.
\end{align*}
In step~4 of the verification procedure, when Trent computes $R_{k_A}\ket{M^2}$, he obtains the correct state except for the first qubit which is now
$$X^aZ^{a\oplus 1}X\ket{M^2_1} = (-1)^{a\oplus 1}X^{a\oplus 1}Z^{a\oplus 1}\ket{M^2_1}.$$
Therefore, as the prefactors only act as an unimportant global phase, the test in step~4 will be passed and give $\theta=1$. The states $\ket{S^0}=\ket{S^1}$ computed by Trent will correspond to the modified state $\ket{R_A}$ and Bob will accept the signature. Then, in a \textsc{Proof-of-Origin} procedure, Bob can prove that Alice sent him the message $\ket{\overline{M}}=X_1\ket{M}$ (i.e., the message $\ket{M}$ to which an $X$ gate has been applied to the first qubit), which constitutes a forgery. Indeed, $\theta=1$ and the test $E_{k_A}(\ket{\overline{R_A}},\ket{\omega_A}) = \ket{S^0}$ in step~4 will pass for the same reason as above. Bob therefore has a forgery attack that works with probability~$1$. This strategy easily generalises: for any operator $U$ that applies a (possibly different) Pauli gate on each qubit, $U=\bigotimes_{i=1}^n \sigma_{\alpha_i}$, Bob can obtain a successful forgery by applying $U$ on $\ket{M^2}$ and on the first $n$ qubits of~$\ket{S}$.

\FloatBarrier
%

\section{Cryptanalysis of a scheme without entangled states}
\label{sec:scheme_no_entangled}

In this section, we review the claims of the AQS scheme of Zhang et al.~\cite{ZXSLL24}, which aims at providing quantum signatures of classical messages without the use of entangled states.

After having recalled the scheme, we review some of the claims made by the authors, among which the claimed impossibility of forgery by Bob, as well as the information-theoretical security of the scheme.

\subsection{Description of the protocol}

\begin{algorithm}[p!]
\begin{algorithmic}
\Procedure{Init}{}
\State 1. $\A\leftrightarrow\T$: $k\in\{0,1\}^\ell$ (where $\ell\gg n$).
\State 2. $\A\leftrightarrow\B$: $x,y\in\{0,1\}^n$.
\EndProcedure{}
\vspace{1mm}

\Procedure{Sign }{\text{on a classical message} $m\in\{0,1\}^\ast$}
\State 1. $\A$: $\ket{M}=F_k(m)$.
\State 2. $\A$: $\ket{S}=\widetilde{Y}^xH^y \ket{M}$.
\State 3. $\A$: $(m,\ket{S})\to\B$ \hfill // secured using decoy states for~$\ket{S}$.
\EndProcedure
\vspace{1mm}

\Procedure{Verify}{}
\State 1. $\B$: $\ket{M'} = H^{y}(-\widetilde{Y})^{x}\,\ket{S}$.
\State 2. $\B$: $(m,\ket{M'})\to\T$ \hfill // secured using decoy states for~$\ket{M'}$.
\State 3. $\T$: $\ket{f'} = (-\widetilde{Y})^{h_k(m)} H^{g_k(m)}\ket{M'}$.
\State 4. $\T$: $f'=$ measure of $\ket{f'}$ according to $\{\ket{0},\ket{1}\}$.
\State 5. $\T$: $f'=f(k\|m)\ ? \ v_T=1$ and stores $(m,f')$ : $v_T=0$; $v_T\to \P_b$.
\State 6. $\B$: $v_T=1$ ? \textbf{accept} : \textbf{reject}.
\EndProcedure

\Procedure{Proof-of-Origin }{\text{on input} $\overline{m}\in\{0,1\}^\ast$}
\State 1. $\T$: $v_T=1$ ? continue : \textbf{favour~$\A$}.
\State 2. $\T$: $(m,f')=(\overline{m},f(k\|\overline{m}))$ ? \textbf{favour~$\B$} : \textbf{favour~$\A$}.
\EndProcedure

\Procedure{Proof-of-Receipt }{\text{on input} $\overline{m}\in\{0,1\}^\ast$}
\State 1. $\T$: $v_T=1$ ? continue : \textbf{favour~$\B$}.
\State 2. $\T$: $m=\overline{m}$ ? \textbf{favour~$\A$} : \textbf{favour~$\B$}.
\EndProcedure

\end{algorithmic}
\caption[AQS Protocol~\ref{alg:AQS_4_32}, of Zhang, Xin, Sun, Li and Li]{AQS protocol of~\cite{ZXSLL24} that does not require entangled states.}\label{alg:AQS_4_32}
\end{algorithm}

In this section, we recall in a synthetic way the protocol of~\cite{ZXSLL24}. One important feature of this protocol is the use of a so-called quantum one way function, denoted $F_k$ for a key $k\in\{0,1\}^\ell$, and which maps classical bit strings to quantum states. It is based on three functions $f,g,h\colon\{0,1\}^\ast \to \{0,1\}^n$, which are used to create $n$-bit strings $f_k(m)=f(k\|m)$, $g_k(m)=g(k\|m)$ and $h_k(m)=h(k\|m)$. The $n$-qubit quantum state created is
\[
F_k(m) = H^{g_k(m)} \widetilde{Y}^{h_k(m)}\ket{f_k(m)}
\]
where $\widetilde{Y}=iY=\left(\begin{smallmatrix}0 & 1 \\ -1 & 0\end{smallmatrix}\right)$. Due to the classical nature of~$f_k(m)$, the effect of $\widetilde{Y}$ is, up to a factorisable phase, the same as~$X$. Therefore, we may write, omitting the unimportant global phase $(-1)^{\sum_i h_i(k\|m) \cdot (1+f_i(k\|m))}$, 
\[
F_k(m) = H^{g_k(m)} X^{h_k(m)}\ket{f_k(m)} = H^{g_k(m)}X^{h_k(m)} X^{f_k(m)} \ket{0} = H^{g_k(m)}X^{\widetilde{f}_k(m)} \ket{0},
\]
where $\widetilde{f}\colon\{0,1\}^\ast\to\{0,1\}^n$ is defined by the bit-wise XOR $\widetilde{f}_k(m)=\widetilde{f}(k\|m)=f_k(m)\oplus h_k(m)$. Similarly, one may also write, omitting a global phase factor $(-1)^{\sum_{i}f_i(k\|m)}$,
\[
F_k(m) = H^{g_k(m)} \widetilde{Y}^{h_k(m)} \widetilde{Y}^{f_k(m)}Z^{f_k(m)}\ket{0} =  H^{g_k(m)}\widetilde{Y}^{\widetilde{f}_k(m)} \ket{0}.
\]
We will use both rewrites during the analysis of the security claims.
The protocol is described in Protocol~\ref{alg:AQS_4_32}.

\paragraph{Ambiguities.} A first ambiguity in~\cite{ZXSLL24} concerns the use of decoy states. It is mentioned that ``sufficient decoy particles" should be used, without specifying how many. It is also mentioned that these decoy particles are inserted randomly in the states, but nothing is said about the distribution of the positions. Moreover, the authors allow the decoy checks to fail up to a certain threshold, but do not explain how to determine it. Also, it is not made explicit what happens if the decoy verification do not succeed. 
In the following, we assume that in the case of a rejection due to a failure of the decoy checks, the parties discard the message and abort, while Trent further stores $v_T=0$.

Another ambiguity concerns the procedures \textsc{Proof-of-Origin} and \textsc{Proof-of-Receipt}. No precise descriptions of them were given in~\cite{ZXSLL24}, but we have inferred them from Section~4.2.3 therein. Another possibility for the \textsc{Proof-of-Receipt} procedure would be to check $(m,f')=(\overline{m},f(k\|\overline{m}))$ in step~2, which would not change the validity of our attacks.

\subsection{Analysis of the security claims}

The authors of~\cite{ZXSLL24} claim that the protocol is secure against existential forgeries and non-repudiation by the sender and the receiver. It is not clear to us whether these security properties are claimed to hold computationally (i.e., for a bounded adversary), or are claimed information-theoretically secure (ITS, i.e., for an unbounded adversary). For instance, Theorem~6 in~\cite{ZXSLL24} involves a random oracle whereas Theorem~3 states that ``the proposed quantum signature has the information-theoretical security". 

Nevertheless, as the scheme can only be used to sign classical messages, its utility would be questionable if it did not provide information-theoretic security. Indeed, classical public-key cryptography (in particular post-quantum cryptography) provides computationally-secure ways of obtaining authenticity and non-repudiation of origin for classical messages that are much easier to implement. We expose here attacks on the non-repudiation and forgery that an adversary with unbounded resources can mount, showing that the scheme is not ITS.

\paragraph{Choice of hash functions and security lemmas.} The protocol does not indicate any restriction on the choice of $f,g$ and~$h$, except that they should be one-way hash functions with uniform output. However, one-wayness is only relevant for computational security. Without further indication, some of the stated results are incorrect, one of which being the Theorem~4 of~\cite{ZXSLL24}, which states that for any message~$m$, the density operator of $F_K(m)$ is uniform (where $K$ is a uniform random variable on $\{0,1\}^\ell$ representing the key~$k$), that is, under proper normalisation, it equals the identity matrix\footnote{While the statement of Theorem~4 is that the output of $F_K(\cdot)$ have same density operator for different input, this is not the property of interest (e.g., a trivial $F_K(\cdot)=0$ would have this feature). Rather, the uniformity of the density operator is what is actually used in the following Theorem~5, hence we discuss it here.}.
As a simple example, let us analyse the choice $f=h$, which could be made for ease of implementation. With such a choice, one has $\widetilde{f}=0$ and the density operator of $F_K(m)$ is $\rho=\bigotimes_{i=1}^n \rho_i$ where
\[
\rho_i = \frac{1}{2} \sum_{g_i\in\{0,1\}} H^{g_i} \ket{0}\bra{0} H^{g_i} = \frac{1}{4} \begin{pmatrix} 3 & 1 \\ 1 & 1\end{pmatrix},
\]
which differs from the identity operator. In the following, we do not expose further the attacks that originate from such degenerate choices, but we stress that one should take care of interactions between primitives when no restrictions or conditions are imposed between each other.

\paragraph{Repudiation and false allegation by Alice with unbounded resources.} If Alice wants to send the message $m$ and wants to be able to repudiate it afterwards, she uses her unbounded resources to find another $m'\in\{0,1\}^\ast$ such that $g_k(m)=g_k(m')$ and $\widetilde{f}_k(m)=\widetilde{f}_k(m')$ (i.e., $f_k(m)\oplus h_k(m)=f_k(m')\oplus h_k(m')$). Such a second preimage exists almost certainly as the message space is infinite ($m\in \{0,1\}^\ast$).

Then, during step~2 of the \textsc{Verify} procedure, she intercepts the message $(m,\ket{M'})$ and replaces it with $(m',\ket{M'})$. Note that this does not interfere with the decoys as Alice only changes the classical part of the communication. Trent's check $f'=f(k\|m')$ will then pass, and he will store $(m',f')$. When Bob wants to prove that he received the message $m$ from Alice, the second check in the \textsc{Proof-of-Origin} procedure will fail and Alice will succeed her repudiation strategy. Moreover, Alice can prove that Bob received $m'$ via a \textsc{Proof-of-Receipt} procedure, which is a false allegation.

This attack can be more sophisticated. For instance, suppose that Alice finds an $m'$ such that the equalities $g_k(m)=g_k(m')$ and $\widetilde{f}_k(m)=\widetilde{f}_k(m')$ hold for every bit except the first one. She can suppose that the first qubit of $\ket{M'}$ is not a decoy state (which has a reasonable success rate, depending on the number of decoy particles). Then, she can substitute $m$ by $m'$ and the first qubit of $\ket{M'}$ by the first qubit of $F_k(m')$ so that Trent's check will pass and $(m',f(k\|m'))$ will be stored, making Alice succeed in her attack.

Notice that, according to~\cite{ZXSLL24}, the key $k$ is shared between Alice and Trent via a quantum key distribution protocol, allowing Alice to have some control over the key~$k$, thus making the task of finding a (quasi-)collision $(m,m')$ easier.

\paragraph{Forgery by Bob with a single signature, for specific hashes.} The first forgery we present uses the supposition that the functions $f$, $g$ and $h$ are of the form
$$f(k\|m)=f_a(k)\oplus f_b(m), \quad g(k\|m)=g_a(k)\oplus g_b(m), \quad h(k\|m)=h_a(k)\oplus h_b(m),$$
with $f_a,f_b,g_a,g_b,h_a$ and $h_b$ chosen independently and at random in the set of functions $\{0,1\}^\ast\to \{0,1\}^n$. Although this assumption might be considered as a very specific choice of hash functions, such choice could be made for performance reasons\footnote{Additionally, we add that a security claim should not depend on unwritten assumptions on the primitives that are used.}. For example, the parties could use the function $f(k\|m) = \text{SHA3}(0\|k)\oplus \text{SHA3}(1\|m)$ in order to pre-compute the hash of a very long key.

We now detail the forgery in the case where the hashes $f$, $g$ and~$h$ (and therefore~$\widetilde{f}$) are of the aforementioned form. Given a message~$m$, one has
\begin{align*}
    \ket{M} &= H^{g(k\|m)}\widetilde{Y}^{\widetilde{f}(k\|m)}\ket{0}\\
            &= H^{g_a(k)\oplus g_b(m)} \widetilde{Y}^{\widetilde{f}_a(k)\oplus \widetilde{f}_b(m)}\ket{0}.
\end{align*}
To forge a valid signature on the message~$\overline{m}$, Bob computes $g_b(m)\oplus g_b(\overline{m})$ and $\widetilde{f}_b(m)\oplus \widetilde{f}_b(\overline{m})$ and modifies the state $\ket{M'}$ computed in step~1 of the \textsc{Verify} procedure to
\begin{align*}
    \ket{\overline{M}} &= H^{g_b(m)\oplus g_b(\overline{m})} \widetilde{Y}^{\widetilde{f}_b(m)\oplus \widetilde{f}_b(\overline{m})}\ket{M'}\\
            &= H^{g_b(m)\oplus g_b(\overline{m})} \widetilde{Y}^{\widetilde{f}_b(m)\oplus \widetilde{f}_b(\overline{m})} H^{g_a(k)\oplus g_b(m)} \widetilde{Y}^{\widetilde{f}_a(k)\oplus \widetilde{f}_b(m)}\ket{0}\\
            &=\pm H^{g_a(k)\oplus g_b(\overline{m})} \widetilde{Y}^{\widetilde{f}_a(k)\oplus \widetilde{f}_b(\overline{m})}\ket{0},
\end{align*} 
where the $\pm 1$ prefactor is the result of the anticommutation relation between $\widetilde{Y}$ and~$H$, $\widetilde{Y}H=-H\widetilde{Y}$, but that only acts as a global phase, hence without any measurable effect on the result. In step~2 of the \textsc{Verify} procedure, Bob sends $(\overline{m},\ket{\overline{M}})$ to Trent. This will pass all the arbitrator's verifications, hence resulting in a successful universal forgery.

\paragraph{Forgery by Bob with polynomially many signatures and unbounded resources.} This final forgery requires adversarial Bob to ask a polynomial number of signatures to the signing party under a reused key~$k$, and assumes that Bob is allowed to launch a \textsc{Verify} procedure for each of them. We note that key reuse is explicitly allowed by the authors of~\cite{ZXSLL24} in their formulation of the security of the scheme against forgery. As in the example provided in~\cite{ZXSLL24}, we fix the parameter $\ell=2n \gg n$.

Bob's strategy consists in modifying the first qubit of $\ket{M'}$ when he sends it to the arbitrator in step~2 of the \textsc{Verify} procedure. Depending on the statistics of the acceptance or rejection of the signature, Bob learns the first bit of the hash functions with high probability. Repeating that process for each qubit, he can learn the whole strings $g_k(m)$ and $\widetilde{f}_k(m)=f_k(m)\oplus h_k(m)$. Using his unbounded resources, Bob can find the key $k$ and forge new messages.

Let us describe this attack in detail for $F_k(m)=\ket{M} = \bigotimes_{i=1}^n \ket{M_i}$. By the construction of~$F_k$, one has $\ket{M_i}= H^{g_i}X^{\widetilde{f}_i} \ket{0}$, where $g_i$ denotes the $i^{\text{th}}$ bit of~$g_k(m)$, and similarly for~$\widetilde{f}_i$. Upon reception of $(m,\ket{S})$ and recovery of~$\ket{M}$, Bob replaces the first qubit of $\ket{M}$ by~$\ket{0}$, and thus sends to the arbitrator the message $(m,\ket{0}\otimes \bigotimes_{i=2}^{n}\ket{M_i})$. If $g_1=0$, Trent will accept or reject with probability~$1$, depending on $\widetilde{f}_1$ being $0$ or~$1$, respectively. If $g_1=1$, Trent will accept or reject with equiprobability. Bob repeats this process $\nu$ times, each requiring a new signature from Alice of the same message. Afterwards, Bob redoes the same scenario but by changing the first qubit by~$\ket{+}$. In that case, for $g_1=1$, the arbitrator accepts or rejects with certainty if $\widetilde{f}_1=0$ or $\widetilde{f}_1=1$ respectively, and his result is unpredictable if $g_1=0$. Table~\ref{tab:results measure Trent} summarises the various cases.

\begin{table}
    \centering
    \renewcommand{\arraystretch}{1.5}
    \begin{tabular}{r|cccc}
    $(g_i, \widetilde{f}_i)$ & $(0,0)$ & $(0,1)$ & $(1,0)$ & $(1,1)$\\
    \hline
    Bob sends $(\bigotimes_{j=1}^{i-1}\ket{M_j})\otimes\ket{0}\otimes (\bigotimes_{j=i+1}^{n}\ket{M_j})$ & $1$ & $0$ & $1/2$ & $1/2$ \\
    Bob sends $(\bigotimes_{j=1}^{i-1}\ket{M_j})\otimes\ket{+}\otimes (\bigotimes_{j=i+1}^{n}\ket{M_j})$ & $1/2$ & $1/2$ & $1$ & $0$ \\
    \end{tabular}
    \caption{Probability of acceptance by the arbitrator of the modified $\ket{M}$ when Bob modifies the $i^{\text{th}}$ qubit of $\ket{M}$ by $\ket{0}$ or~$\ket{+}$, depending on the bits $g_i$ and~$\widetilde{f}_i$.}
    \label{tab:results measure Trent}
\end{table}

After $2\nu$ messages, there is a probability $\frac{1}{2^{\nu-1}}$ that Bob cannot exactly determine the bits $g_1$ and~$\widetilde{f}_1$. Indeed, the only way Bob will not be able to certainly determine $g_1$ and $\widetilde{f}_1$ is when the events with $1/2$ probability of acceptance all give the same output. This occurs with probability $2^{-(\nu-1)}$, and in this case, Bob can guess the correct values of $g_1$ and $\widetilde{f}_1$ with $\frac{1}{2}$ probability. In other words, Bob knows the correct bits with probability $1-\frac{1}{2^\nu}$.
By doing this strategy for all qubits, Bob learns the values of $g(k\|m)$ and $\widetilde{f}(k\|m)$ with probability $(1-2^{-\nu})^n$. Choosing $\nu=n$, this probability tends to $1$ for large messages, which is the case studied by the authors ($n=128$). This adversarial strategy only requires a polynomial amount of messages as it requires at most $2\nu\cdot n=2n^2$ signatures.

With the knowledge of $g(k\|m)$ and $\widetilde{f}(k\|m)$, the adversarial Bob, with unbounded computing capacities, can obtain the key $k$ with large probability. Indeed, by uniformity, there are $\frac{2^{\ell+|m|}}{2^n}$ preimages of $g(k\|m)$, among which only about $2^{\ell-n}$ end with the correct message~$m$. Each of these preimages for $g$ gives the correct $\widetilde{f}(k\|m)$ with probability~$2^{-n}$. Hence, with unbounded resources, Bob finds the key $k$ with large probability (as $\ell=2n$), and can mount any forgery of his choice\footnote{To phrase it differently, both $g(k\|m)$ and $\widetilde{f}(k\|m)$ give Bob $n$ bits of information on the key, which is of size $\ell=2n$.}. 

This attack also works for longer keys whose size is polynomial in the message length. Indeed, Bob extracts $2n$ bits of information in $2n^2$ signatures of a fixed message, and can iterate this process with different messages, hence obtaining the key in a polynomial amount of queries (with an exponentially low probability of failure).

\paragraph{Disavowal by Bob.} If Bob receives a pair $(m,\ket{S})$ that he dislikes, he can simply discard it (or retain it for later). This issue, due to the message $m$ being fully available to Bob at the end of the \textsc{Sign} procedure, has already been discussed in Subsections~\ref{subsec:analyse_2} and~\ref{subsec:analyse_3}. Note that many of the recent AQS schemes delay the full disclosure of the message to the recipient until the last step of verification, after Bob has communicated with the arbitrator (e.g., \cite{ZQ10,LS15,ZK20}).

Even if the trivial disavowal strategy is forbidden, a corollary of the forgery detailed above is that a malicious Bob with unbounded resources can disavow messages. Indeed, once he has obtained the key~$k$, Bob can modify the message to his liking and forge its signature so that Trent publishes $v_T=1$. A \textsc{Proof-of-receipt} started by Alice with her message will fail, and Bob will have succeeded in his repudiation. 

\section{Conclusion}

In this article, we have shown that several recent AQS schemes, designed to provide secure signatures with an arbitrator, suffer from security flaws. In many cases, the core AQS security features (unforgeability and the impossibility of disavowal by either sender or receiver) are not met. Additionally, we have often identified specification ambiguities that render the claimed guarantees difficult to verify. These gaps in the specification may even lead to unforeseen attacks. This enlightens one more time the necessity of precise descriptions of algorithms and protocols, precise security definitions and comprehensive proofs thereof.

\FloatBarrier


\end{document}